\documentclass{nature}
\usepackage{graphicx}
\makeatletter
\let\saved@includegraphics\includegraphics
\AtBeginDocument{\let\includegraphics\saved@includegraphics}
\renewenvironment*{figure}{\@float{figure}}{\end@float}
\makeatother

\usepackage{color}
\usepackage{graphicx}
\usepackage{epstopdf}
\usepackage{bm}
\usepackage{dcolumn}
\usepackage{verbatim}
\usepackage{mathbbol}
\usepackage{bm}
\usepackage{dcolumn}
\usepackage{amsmath}
\usepackage{placeins}
\usepackage{mathrsfs}
\usepackage{gensymb}
\usepackage{physics}
\usepackage{upgreek}
\usepackage{subcaption}
\usepackage{MnSymbol}
\usepackage{float}
\usepackage[colorinlistoftodos]{todonotes}

\usepackage[resetlabels,labeled]{multibib}

\newcites{M}{References}

\usepackage[normalem]{ulem}

\newcommand{\approptoinn}[2]{\mathrel{\vcenter{
  \offinterlineskip\halign{\hfil$##$\cr
    #1\propto\cr\noalign{\kern2pt}#1\sim\cr\noalign{\kern-2pt}}}}}

\title{Topological strong field physics on sub-laser cycle time scale}

\author{R. E. F. Silva$^{1,2,\dagger}$, \'A. Jim\'enez-Gal\'an$^{1,\dagger\dagger}$, B. Amorim$^{3}$, O. Smirnova$^{1,4}$, \& M. Ivanov$^{1,5,6}$}

\begin{document}

\maketitle

\begin{affiliations}
 \item Max-Born-Institute, Max-Born Stra{\ss}e 2A, D-12489 Berlin, Germany.
 \item Department of Theoretical Condensed Matter Physics, Universidad Aut\'onoma de Madrid, E-28049 Madrid, Spain
 \item CeFEMA, Instituto Superior T\'ecnico, Universidade de Lisboa, Av. Rovisco Pais, 1049-001 Lisboa, Portugal
 \item Technische Universit\"at Berlin, Ernst-Ruska-Geb\"aude, Hardenbergstra{\ss}e 36A, 10623 Berlin, Germany.
 \item Department of Physics, Humboldt University, Newtonstra{\ss}e 15, D-12489 Berlin, Germany.
 \item Blackett Laboratory, Imperial College London, South Kensington Campus, SW7 2AZ London, United Kingdom.
\end{affiliations}
\noindent $\dagger$ silva@mbi-berlin.de
\noindent $\dagger\dagger$ jimenez@mbi-berlin.de


\begin{abstract}

Sub-laser cycle time scale of electronic response to strong laser fields enables
 attosecond dynamical imaging in atoms, molecules and solids~\cite{Krausz2009, Baker2006, Shafir:2012aa,Hohenleutner:2015aa}. 
Optical tunneling and high harmonic generation~\cite{Baker2006,Smirnova:2009aa,Eckart:2018aa,Ghimire:2010aa} are the hallmarks of attosecond
 imaging in optical domain, including imaging of phase transitions in solids~\cite{Silva:2018aa,Bauer:2018aa}. 
Topological phase transition yields a state of matter intimately linked with electron dynamics, 
as manifested via the chiral edge currents in topological insulators~\cite{Hasan:2010aa}. Does topological state of
 matter leave its mark on optical tunnelling and sub-cycle electronic response?
 We identify distinct topological effects on the directionality and the attosecond timing
 of currents arising during electron injection into conduction bands.
We show that electrons tunnel across the band gap differently in trivial and topological phases, 
for the same band structure, and identify the key role of the Berry curvature in this process.
 These effects map onto topologically-dependent attosecond delays in high harmonic emission and
 the helicities of the emitted harmonics, which can record the phase diagram of the system and
 its topological invariants.  Thus,  the topological state of the system controls its attosecond,
 highly non-equilibrium electronic response to strong low-frequency laser fields, in bulk. 
Our findings create new roadmaps in studies of topological systems, 
building on ubiquitous properties of sub-laser cycle strong field response - a unique mark of attosecond science.

\end{abstract}



Intense light  incident on a quantum system can drive the system's electrons far from their 
equilibrium within a fraction of its cycle. 
Accelerated by the field and by the forces
inside the system, the electrons emit coherent radiation which 
can contain 
many tens (or hundreds) of harmonics of the incident light~\cite{Krausz2009}. 
Such ultra-broad, coherent spectrum 
implies access to extreme time resolution of the underlying  charge dynamics. It is the basis 
of high harmonic generation spectroscopy 
\cite{Baker2006, Smirnova:2009aa, Shafir:2012aa}, now 
rapidly expanding beyond atomic and molecular physics 
towards  characterizing fundamental ultrafast processes in solids. 
Examples include the observation of dynamical Bloch oscillations~\cite{Ghimire:2010aa, Schubert:2014aa,Luu:2015aa}, 
band structure tomography~\cite{Vampa:2015aa,Tancogne-Dejean:2017aa},  
resolving electron-hole dynamics~\cite{McDonald:2015aa, Schubert:2014aa, Bauer:2018aa} and
light-driven phase transitions in a Mott insulator~\cite{Silva:2018aa}, the Peierls
phase transition~\cite{Bauer:2018aa}, or the imprint of the Berry phase on 
optical response~\cite{Liu:2016aa,Luu:2018aa}.  
Complete characterization of the emitted light, both amplitude and spectral phase, 
allows one to recover time-frequency maps of the emission and thus 
resolve the underlying charge dynamics~\cite{Shafir:2012aa, Hohenleutner:2015aa}. 
Here, we apply these ideas to a topological insulator.


The discovery of the integer quantum Hall effect (IQHE) and its subsequent link 
to the so-called topological invariants of the system's bulk~\cite{Thouless:1982aa}, have led to the discovery of
topological phases of matter.   In a topological insulator, the topological phase supports states  that carry currents around 
the insulator's edges. ``Protected'' by the topological 
invariants of the bulk, the chiral edge states are robust to perturbations, making them appealing for applications, 
for example in dissipationless devices or in topologically robust superconductors~\cite{Hasan:2010aa}.
In two-dimensional (2D) materials exhibiting 
the IQHE, the topological invariant is the Chern number -- 
the integral of the Berry curvature over the 2D-Brillouin zone of the filled bands.

\begin{figure}
\includegraphics[width=\linewidth]{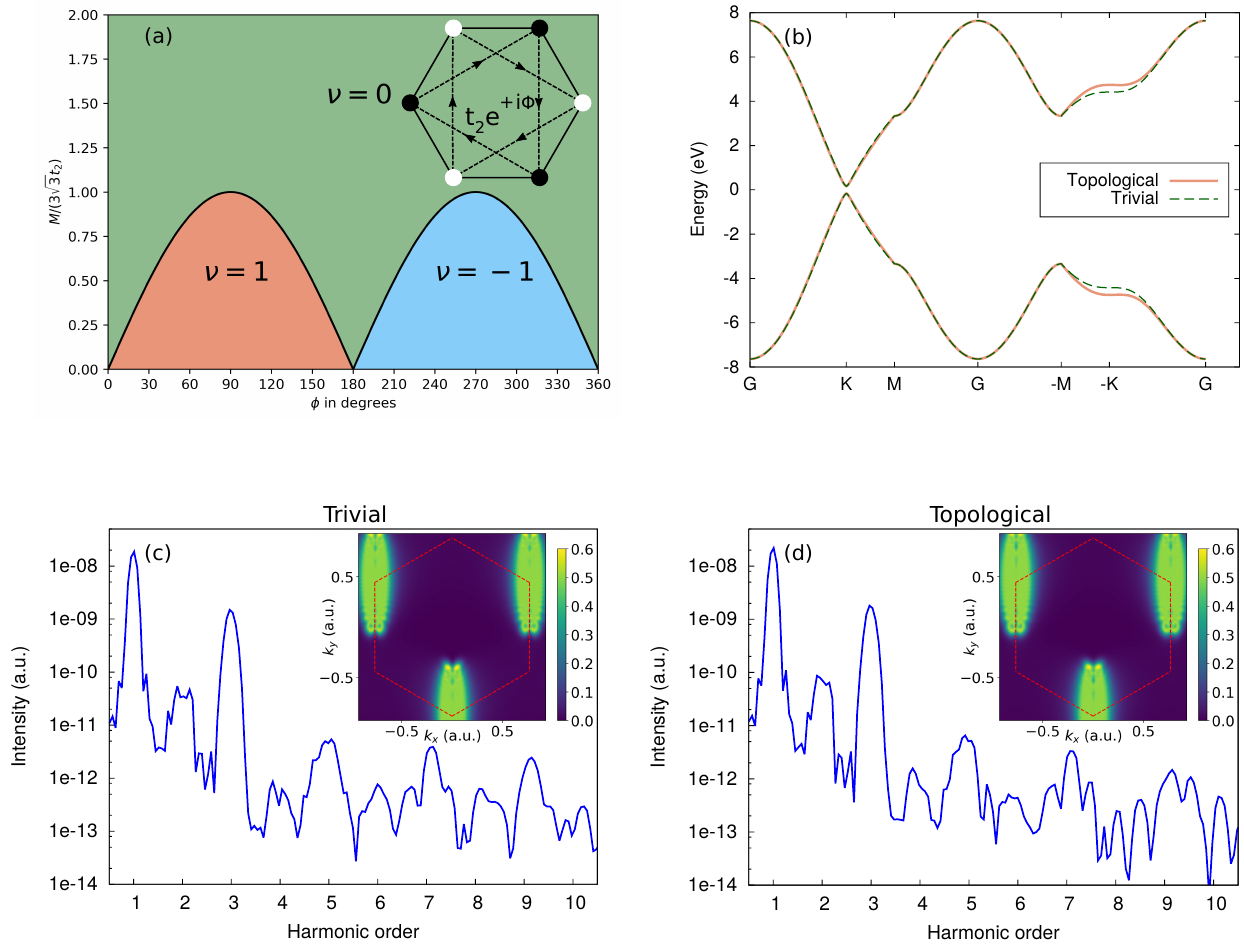}
\caption{\textbf{The Haldane system in two phases}. (a) The phase diagram: $\nu=0$ -- trivial phase, $\nu=\pm 1$ -- topological phase. 
Inset shows the real-space hexagonal lattice with two different atoms (black and white, on-site energies $\pm M$). 
Arrow lines show the second-neighbour hopping $t_2 e^{i\phi}$ with positive phase (negative phase is reversed).
(b) Bulk band structures before and after the phase transition are very similar
($M/(3\sqrt{3}\,t_2) = 1.07$, green dashed, and $M/(3\sqrt{3}\,t_2) = 0.93$, orange solid).
(d,e) High harmonic spectra and electron densities at the end of the pulse (as insets) at $M/(3\sqrt{3}\, t_2) = 1.07$ (trivial phase, (c)) and $M/(3\sqrt{3}\,t_2) = 0.93$ (topological phase, (d)) do not
readily distinguish the two phases.}
\label{fig:Haldane}
\end{figure} 
Remarkably, in spite of extensive work on topological materials, their associated ultrafast 
dynamics has been hardly explored, with very few exceptions~\cite{Bauer:2018aa,Reimann:2018aa}. Here are some of the key unanswered questions:
Does the 
highly non-equilibrium electron dynamics in the bulk, 
driven by a strong laser field, encode the topological properties on the sub-laser cycle
time-scale? How do the Berry curvature 
and the Chern number affect the first step in the nonlinear response --
the field-driven injection of electrons across the bandgap? 
We answer these questions using the paradigmatic example of the topological insulator, the
Haldane system \cite{Haldane:1988aa} (Fig.~\ref{fig:Haldane}). 
We show how its topological properties affect the geometry, directionality, and 
the attosecond timing of the injected currents. We further show that high harmonic emission
encodes topologically-dependent attosecond delays induced in these currents, and 
that the helicities of the emitted harmonics encode the phase state of the system and its topological invariants. 
Attosecond characterization of topological effects on light-driven electron currents is a crucial 
step towards petahertz electronics in topological materials~\cite{Garg:2016aa}.

In the Haldane system, the hexagonal lattice hosts different atoms, A and B, on the adjacent
sites. Their different on-site energies $\pm M$  open 
the bandgap at the points $\mathbf{K}$ and $\mathbf{-K}$ of the
Brillouin zone. The complex-valued 
second-next neighbour  hopping $t_2 e^{i\phi}$ (Fig.~\ref{fig:Haldane}a) controls the 
topological state. 
The system is a trivial insulator with zero Chern number, $\nu=0$, if $|M/(3\sqrt{3})| > t_2 \sin \phi$, but 
becomes topological when  $|M/(3\sqrt{3})| < t_2 \sin \phi$ (Fig.~\ref{fig:Haldane}b), with $\nu=\pm 1$, 
giving rise to a non-zero Hall conductivity and the appearance of gapless edge states. 
This model,  recently demonstrated experimentally~\cite{Jotzu:2014aa},
is a gateway for studying topological properties of materials such as the quantum spin Hall 
effect~\cite{Kane:2005aa} or the valley Hall effect~\cite{Mak2014}.  

%
%

Strong field response in solids naturally emphasizes the role of the
material band structure~\cite{Tancogne-Dejean:2017aa,Ghimire:2010aa,Vampa:2017aa,Ndabashimiye:2016aa,Hawkins:2015aa}.
Indeed, optical tunnelling is exponentially sensitive to the bandgap, while 
the band structure determines the subsequent motion of electrons and holes. 
The transition between the trivial and the topological  phases 
is indeed accompanied by marked changes in the bulk band structure: the gap between 
the valence and the conduction bands closes and then re-opens during the transition (Fig.~\ref{fig:Haldane}c). 
High harmonic spectra will inevitably reflect these changes \cite{Chacon2018}.

Alas, they do not readily distinguish the two phases of the material.  
Figs.~\ref{fig:Haldane}(c,d)
show high harmonic spectra generated before and after the phase transition, using intense 
low-frequency field polarized linearly in the y-direction
(field wavelength $\lambda=3\mu$m,  peak amplitude $E_0=4\times 10^9$ V/m). 
The system at $t_2 = 0$ has a bandgap of 4.6 eV, chosen similar to a single layer of hexagonal boron nitride within
the two band model. 
The band structure is very similar before and after the transition (Fig.~\ref{fig:Haldane}b), and so is the 
harmonic spectrum. 
Attributing specific spectral features to the Berry curvature and the Chern number
is hardly straightforward. 

However, analogies between strong field response in atoms, 
molecules and solids ~\cite{Vampa:2017aa} suggest that 
other key aspects of electronic structure, besides the bandstructure, 
should play an important role. In atoms and molecules, these are
the angular momentum and the chirality of the 
states  ~\cite{Barth:2011aa,  Cireasa:2015aa, Eckart:2018aa}. 
We shall see how the Berry curvature takes similar role in strong field response in solids, mapping 
the topological state of the system onto the timing and the directionality of the injected 
currents. The currents control harmonic polarization and temporal structure. 

\begin{figure}
\includegraphics[width=\linewidth]{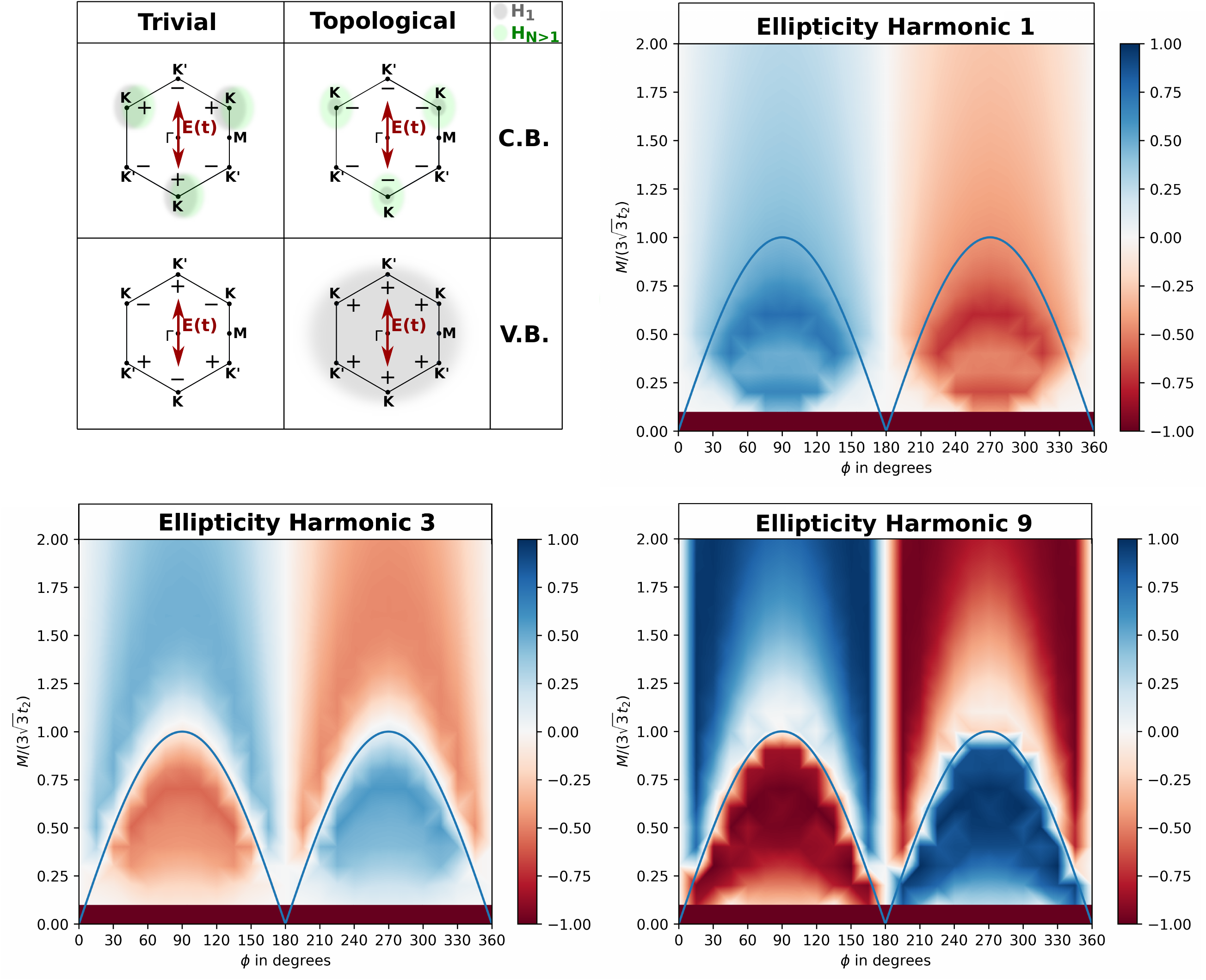}
\caption{\textbf{Ellipticity of harmonics in topological insulators}. Top left: Scheme of the regions in the Brillouin zone probed by the first harmonic (grey shadow) and odd high harmonics (green shadow), see text for details.  Ellipticity 
of the first (top right), third (bottom left) and ninth (bottom right) harmonics as a function of the amplitude $t_2$ and phase $\phi$ of the next-nearest neighbour hopping. The blue line indicates the topological phase transition.}\label{fig:ellipticity}
\end{figure}

Figure~\ref{fig:ellipticity} confirms these expectations. In contrast to the harmonic spectra, 
the phase transition is clearly recorded in the helicities of the emitted harmonics. The 
helicity of the linear response (H1, first harmonic) 
provides a reference, while the helicities of higher harmonics (here we show 
H3 and H9) switch exactly at the phase transition. Tracking the  
helicity of H3 relative to H1 generates the phase diagram of the system 
(see supplementary note 1).


The key physics  can be understood semiclassically.
Irrespective of the specific mechanism responsible for the 
harmonic emission (intraband current or 
electron-hole recombination), harmonics are associated with
the injection of charge across the bandgap.
As the laser field (vector-potential 
$\mathbf{A}_L(t)$, field $\mathbf{E}(t)=-\partial \mathbf{A}_L/\partial t$, linearly polarized 
along the $y$ axis) drives the injected electrons, changing the initial quasimomentum ${\bf k}_i$ to
$\mathbf{k}=\mathbf{k}(t)=\mathbf{k}_i+ e \mathbf{A}_L(t)$, the non-zero Berry curvature 
$\mathbf{\Omega} (\mathbf{k})$ generates orthogonal motion in the $x$-direction, 
\begin{equation}
\label{eq:semiclassical_maintext}
\dot{\mathbf{r}}_n(t) = \nabla_{\mathbf{k}}\,\varepsilon_n (\mathbf{k}(t)) + e 
\mathbf{E}(t) \times \mathbf{\Omega}_{n} (\mathbf{k}(t)).
\end{equation}
Here, $n$ labels the band,
$\varepsilon_n (\mathbf{k})$ is the band dispersion.
The motion orthogonal to the field (along $\bf x$) means that the emitted harmonics acquire the $\bf x$ component. 
Qualitatively, 
$\dot{\textbf{r}}_{n}(t)\cdot \hat{e}_y$ follows the vector-potential $\textbf{A}_L(t)$  and 
$\dot{\textbf{r}}_{n}(t)\cdot \hat{e}_x$ follows the field $\textbf{E}(t)$. Hence, the $\bf x$ and $\bf y$ harmonic 
components are $\pi/2$ out of phase, i.e. the harmonics are elliptically polarized (see Methods for a 
rigorous discussion). 

As expected, in the trivial phase the helicities
of all harmonics are the same. As  
the Berry curvature changes sign upon the topological transition, 
the anomalous current in the conduction band changes direction. Hence,
the harmonics will also flip their helicity. In contrast, the 
response at the fundamental frequency in the topological 
phase does not change the helicity. The reason is that it is now 
dominated by  the  
non-zero Hall conductivity of the valence band,  
where the Berry curvature is opposite to that in the conduction band and where most of the electrons are. Thus, the helicity of H1 
is  naturally opposite to that of H3 (or H9), as Fig.\ref{fig:ellipticity} shows (see Methods for further details). 


Our previous analysis has tacitly implied that, for similar bandstructures
in the trivial and topological phases, the time-dependent electron densities 
promoted to the conduction band  
would also be similar. This is indeed
a good first approximation, see insets in Fig. \ref{fig:Haldane}(c,d). As expected, the injection occurs near the peaks of the oscillating electric field (${\bf A}_L(t)=0$) and near  
the minima of the bandgap,
i.e. in the vicinity of the $\bf K$ points of the Brillouin zone.
Once injected, the electron's quasi-momenta cluster around 
${\bf k}(t)={\bf K}+ e{\bf A}_L(t)$. 

\begin{figure}
\includegraphics[width=\linewidth]{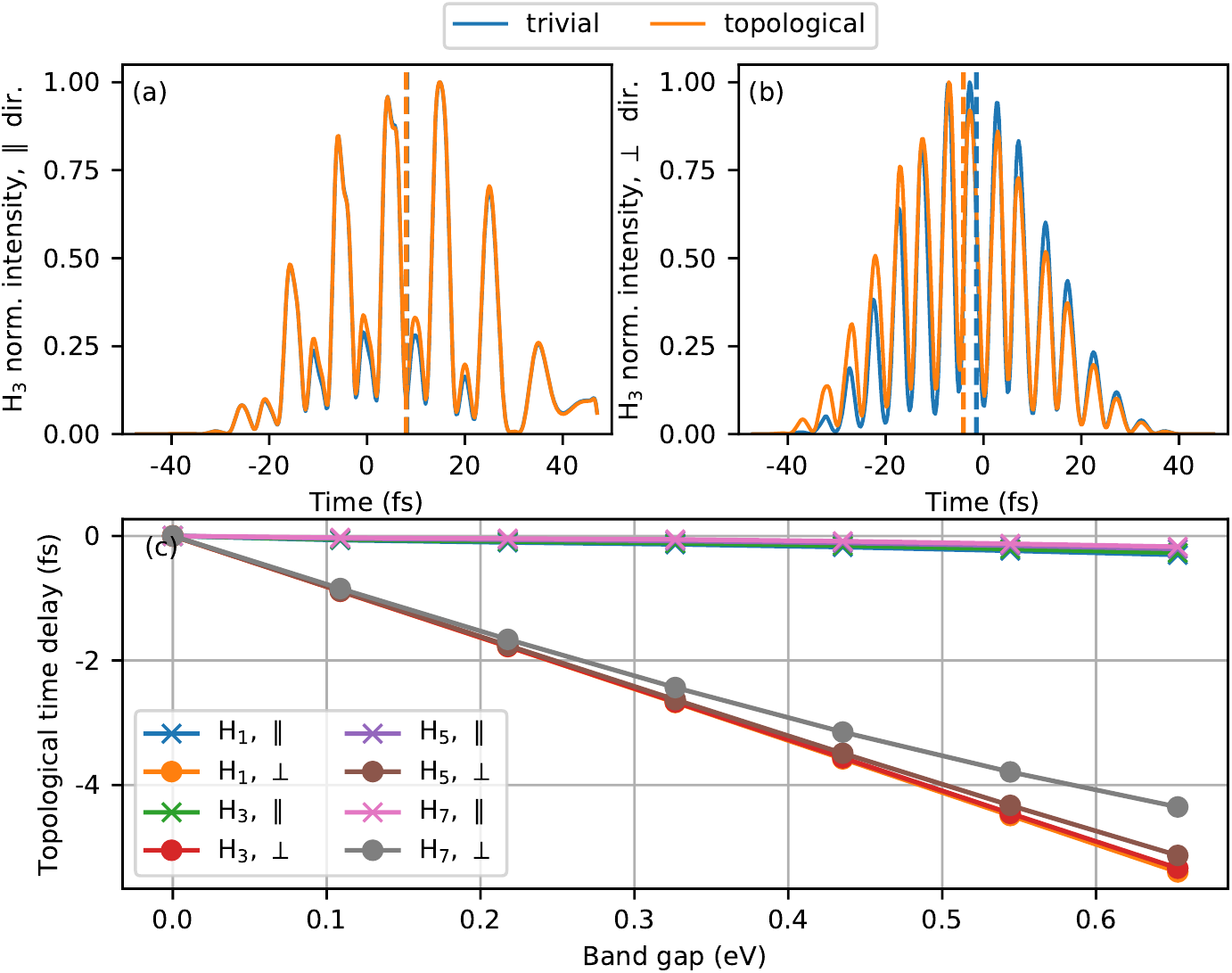}
\caption{\textbf{Topological time delays in harmonic emission}. Time-frequency 
maps of the emission in the two
phases shown in Fig.~\ref{fig:Haldane}b almost overlap for the parallel component (a) but
show a clear asymmetry between positive and negative times for the anomalous component (b), yielding topologically-dependent
time-delays, $\tau$ = $\langle \tau \rangle_{\text{topo}} - \langle \tau \rangle_{\text{trivial}}$, in the emission of a given harmonic (c), which vary with the bandgap. The emission time delay is defined as the weighted average, $\langle \tau \rangle = \frac{\int f(t)\,t\,dt}{\int f(t)\,dt}$, where $f(t)$ are the frequency-resolved time maps shown in panels (a) and (b). Panels (a) and (b) show harmonic 3 for a 0.32 eV band gap, with the blue and orange vertical dashed lines indicating the weighted average times of ionization in the trivial and topological phases, respectively.}
\label{fig:time_delays}
\end{figure}

However, our analytical analysis 
of the injection step, provided in the Methods section, demonstrates  that
the injected densities are different in the trivial and topological
phases, for the same bandstructure. Specifically, the optimal
$\bf k$ for tunnelling across the bandgap, which maximizes the 
injection rate, is shifted from the 
${\bf K}$-point. In the parabolic band approximation 
near the ${\bf K}$-point, the shift orthogonal to the driving field is
\begin{equation}
\label{eq:anomalous_shift}
\Delta\mathbf{k}_{\perp}\simeq m_{\perp} \mathbf{E}(t) \times 
\left[
\mathbf{\Omega}_{c} (\mathbf{k^{(0)}_{\parallel},k_{\perp}^{(0)}})-
\mathbf{\Omega}_{v} (\mathbf{k^{(0)}_{\parallel},k_{\perp}^{(0)}})
\right].
\end{equation}
Here ${\bf k}^{(0)}_{\parallel},{\bf k}_{\perp}^{(0)}$ are the optimal values 
in the absence of the Berry curvature, $\parallel$ and $\perp$ stand for  'parallel' and 'perpendicular' with respect to the laser polarization,
$\mathbf{\Omega}_{v}, \mathbf{\Omega}_{c}$ are the Berry curvatures in the 
valence and conduction bands, and $m_{\perp}$ is the electron-hole effective reduced  mass (Eq.(\ref{eq:anomalous_shift})
assumes that the band gap is above the driving laser frequency).
Note that $\Delta{\bf k}_{\perp}$ has opposite sign in the trivial and the topological phases. 
Non-zero $\Delta{\bf k}_{\perp}$ 
means that, for non-zero Berry curvature,  
the anomalous currents reach the same region 
of the BZ and the same velocities at different times, shifted by $\tau \propto \Delta{\bf k}_{\perp}/\mathbf{E}$.  
Hence, irrespective of the harmonic emission mechanism 
(current-- or recombination-driven), there should be a  Berry curvature-induced  
time-delay in the emission of a given harmonic between the trivial and the topological phases, 
for the harmonic component orthogonal
to the driving field. This delay should increase as we move away from the transition point, reflecting
the increasing effective mass $m_{\perp}$.

This is confirmed by our numerical simulations.  Fig. \ref{fig:time_delays}(a,b) show the 
time-frequency analysis of high harmonic emission in the trivial and topological phases 
(here only H3 is shown, see supplementary note 2 for other high harmonics). We
made sure to compare cases with virtually 
the same band structures in both phases (see Fig. \ref{fig:Haldane}(b) and supplementary note 2).  For 
the harmonic polarization parallel  to the driving field,  the 
frequency-resolved time maps of the emission in the two
phases almost overlap, signaling a negligible influence of the band structure. For the 
orthogonal polarization component, they visibly do not, yielding time-delays in the emission 
of a given harmonic exclusively linked to the topological properties of the material, Fig. \ref{fig:time_delays}(c). 

\begin{figure}
\includegraphics[width=\linewidth]{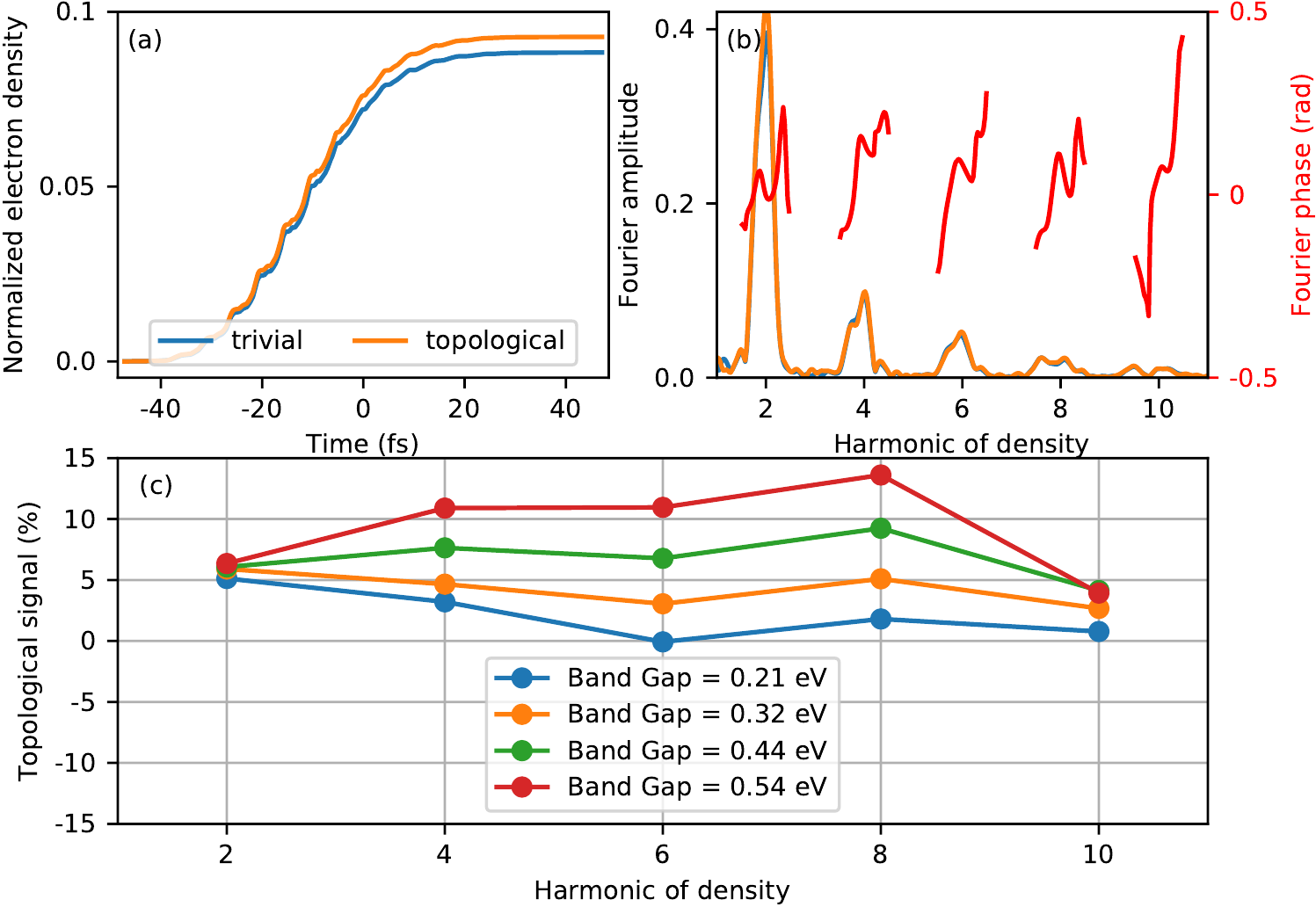}
\caption{\textbf{Topological signal in injected currents}. Total injected electron density (a) and
its Fourier transform (b), for the trivial (blue) and the topological (orange) phases with the band structures
in Fig.~\ref{fig:Haldane}b (band gap = 0.32 eV). The Fourier phase (panel b, red line) refers to the difference in the Fourier phases between the topological and trivial systems. Panel (c) shows the topological signal in the harmonic amplitude of the electron density for different band gaps.}
\label{fig:topo_dichroism}
\end{figure} 

Figs.\ref{fig:topo_dichroism}(a,b) show the total injected electron density integrated over the whole BZ and its Fourier transform, $\rho_{cc} (N\omega)$, 
 for the trivial and the topological phases. It directly confirms the dependence of the injected 
 electron density on the Berry curvature, for the same band structure. 
For high harmonics of the density, the topological signal, defined in the frequency domain as 
\begin{equation}
\label{eq:topological_dichroism}
{\rm TS}_{\rho}(N\omega)=2\frac{\left[|\rho_{cc,top}(N\omega)|-|\rho_{cc,triv}(N\omega)|\right]}
{|\left[\rho_{cc,top}(N\omega)|+|\rho_{cc,triv}(N\omega)|\right]}
\end{equation}
grows with the bandgap, and, for the cases studied in this work, reaches values of 15\%.

The topological dependence in
the time-dependent density of the injected charges can also manifest in other observables, for example 
in the helicity maps of the emitted harmonics or in sudden changes in the 
harmonic phases (see supplementary note 3).
Let us write the momentum-resolved electron density in the conduction band $\rho_{cc}({\bf k},t)$ as
\begin{equation}
\label{eq:Berry_density}
\rho_{cc}({\bf k},t)=
\rho_{cc}^{(0)}({\bf k},t; \Omega=0)+\Delta \rho_{cc}(k,t; \Omega)\simeq 
\rho_{cc}^{(0)}({\bf k},t)+\Omega({\bf k})\frac{\partial \rho_{cc}}{\partial \Omega}|_{\Omega=0} (k,t),
\end{equation}
where $\rho_{cc}^{(0)}({\bf k},t)$ is the Berry curvature-independent injected density, 
and the second term is the correction, linear in the Berry curvature in the first order.  While 
this additional term is small (Fig. \ref{fig:Haldane}(c,d)), its 
contribution to the anomalous current is quadratic in $\Omega$ and does not change sign during the phase transition, in contrast to the main
contribution. If the main contribution to the harmonic emission, associated with 
$\rho_{cc}^{(0)}({\bf k},t)$, is suppressed for a particular harmonic, the additional contribution  
associated with $\Delta \rho$ could become visible and the helicity of the harmonic
should flip. We observe such flips in the helicity map of harmonic 5 (see supplementary note 1).
This is accompanied by a sudden spectral phase jump in the harmonic line, reflecting destructive
interference in the main contribution to the 
harmonic emission (see supplementary note 3).

Thus, the topological properties of materials manifest 
in the polarization and timing of the emitted
harmonics. Analysis of these features allows all-optical retrieval of the complete 
phase diagram of  the paradigmatic Chern insulator, the Haldane system, 
and identifies time-delays in the harmonic emission associated with
the Berry curvature.  
Our work introduces strong-field and attosecond physics to topological 
materials, with high harmonic spectroscopy as a tool for characterizing topological invariants 
in the condensed phase, and for time resolving the ultrafast topological 
nonlinear response arising upon the phase transition.

\section*{References}
\bibliographystyle{naturemag}
\bibliography{biblio}

\subsection{Data availability.} The data that support the findings of this study are available from the corresponding author upon request.

%

\begin{addendum}
\item R.E.F.S. and M.I. acknowledge support from EPSRC/DSTL MURI grant EP/N018680/1. A.J.G and M.I. acknowledge support from the DFG QUTIF grant IV 152/6-1. O.S. acknowledge support from the DFG SPP 1840 ``Quantum Dynamics in Tailored Intense Fields'' project SM 292/5-1; and  MEDEA project, which has received funding from the European Union's Horizon 2020 research and innovation programme under the Marie Sk\l{}odowska-Curie grant agreement No 641789. B.A. received funding from the European Union's Horizon 2020 research and innovation programme under grant agreement No. 706538.
\end{addendum}

\end{document}